\documentclass{desyproc}
\usepackage{aas_macros}
\newcommand{\jcap}{JCAP}
\begin{document}

\title{Modelling $\gamma$-ray-axion-like particle oscillations in turbulent magnetic fields: relevance for observations with Cherenkov telescopes}

\author{{\slshape Manuel Meyer}\\[1ex]
The Oskar Klein Center for CosmoParticle Physics, Department of Physics, Stockholm University, Albanova, SE-10691 Stockholm, Sweden}

\contribID{Meyer\_Meyer}

\desyproc{DESY-PROC-2014-XX}
\acronym{Patras 2014} 
\doi  

\maketitle

\begin{abstract}
Axion-like particles (ALPs) are a common prediction of certain theories beyond the Standard Model and couple to photons in 
the presence of external magnetic fields.  
As a consequence, photon-ALP conversions could lead to an enhancement of the flux of extragalactic $\gamma$-ray sources
that is otherwise attenuated due to the interactions with background radiation fields.
The magnetic fields traversed by the $\gamma$ rays are often turbulent and frequently modelled with a simple domain-like structure. 
Given a maximum mixing between photons and ALPs, we show that in such models realisations of the fields exist 
for which the photon-ALP oscillation probability vanishes.
This behaviour does not occur in more sophisticated magnetic-field models.

\end{abstract}
\section{Introduction}
Very high energy $\gamma$ rays (VHE; energy $E \gtrsim100\,$GeV) originating from extragalactic objects interact with 
photons of the extragalactic background light (EBL) leading to an exponential attenuation of the $\gamma$-ray flux emitted by the source~\cite{dwek2013}.
Direct detections of the EBL are difficult due to foreground emission~\cite{hauser1998} and thus the exact level of the EBL photon density remains unknown.
Recent EBL models (e.g. Refs.~\cite{kneiske2010,dominguez2011,gilmore2012}) predict densities close to lower limits deduced from galaxy number counts~\cite{madau2000,fazio2004}.
In simple emission models of blazars\footnote{Blazars are active galactic nuclei (AGN) with a jet closely aligned to the line of sight. It is the most common source class 
for extragalactic VHE $\gamma$-ray emitters, see e.g. \url{http://tevcat.uchicago.edu/}.},
no spectral hardening is expected at VHE and consequently spectra corrected for EBL absorption should not show such features. 
Nevertheless, evidence for such signatures has been found~\cite{deangelis2009,deangelis2011,dominguez2011alps,horns2012,rubtsov2014}. 
An explanation might be the oscillations of $\gamma$-rays into axion-like particles (ALPs), 
spin-0 pseudo-Nambu-Goldstone bosons that arise in certain Standard Model extensions.
These particles couple to photons in external magnetic fields (see Ref.~\cite{jaeckel2010} for a review)
and could lead to a flux enhancement as ALPs do not interact with EBL photons.

Several turbulent magnetic-field environments have been studied in this respect including the intergalactic magnetic field~\cite{mirizzi2007,deangelis2007}, the AGN host galaxy~\cite{tavecchio2012},
intra-galaxy-cluster fields~\cite{horns2012ICM}, magnetic fields in AGN lobes~\cite{tavecchio2014}, and in the Milky Way
\cite{simet2008}. 
The turbulent fields are commonly modelled with a simple domain-like structure: the path length is split up into $N$ domains of coherence length $L_\mathrm{coh}$. 
While the field strength remains constant over all cells, the orientation of the field is assumed to change randomly from one cell to the next. 
The random nature of the field makes it necessary to calculate the conversion probability for a large number of random realisations. 
As $N$ grows, the photon-ALP oscillations can often only be calculated numerically. 
Here, we show analytically that in these simple models realisations exist for which the photon-ALP conversion probability vanishes. 

\section{Photon-ALP oscillations}
The equations of motion of a monochromatic photon-ALP beam composed of the two photon polarisation states $A_{1,2}$ and the ALP field strength $a$, $\Psi = (A_1,A_2,a)^T$,
of energy $E$ propagating along the $x_3$ axis in a cold plasma with homogeneous magnetic field 
can be written as~\cite{raffelt1988}
\begin{equation}
\left(i\frac{\mathrm{d}}{\mathrm{d}x_3} + E + \mathcal{M}_0\right)\Psi(x_3) = 0,
\label{eq:eom}
\end{equation}
where the mixing is induced by off-diagonal 
elements of the mixing matrix $\mathcal M_0$. 
The resulting photon-ALP oscillations are similar to neutrino oscillations and we denote the mixing angle by $\alpha$
 (see, e.g. Ref.~\cite{bassan2010} for the full expressions for $\mathcal{M}_0$ and $\alpha$).
Equation \eqref{eq:eom} can be solved with the transfer matrix
$\mathcal T$, so that  $\Psi(x_3) = \mathcal{T}(x_3,0;\psi;E)\Psi(0)$, where $\psi$ denotes the angle between the transversal magnetic field and the photon polarisation state along $x_2$~\cite{deangelis2011}.
 With the eigenvalues $\lambda_j$, $j = 1,2,3$, of the mixing matrix  and introducing the notation $s_\theta = \sin\theta$ and $c_\theta = \cos\theta$, 
the transfer matrix can be written as~\cite{deangelis2011}
\begin{eqnarray}
\mathcal{T} = \sum\limits_{j = 1}^3 e^{i\lambda_j x_3} \mathcal{T}_j \qquad\mathrm{with}&
\mathcal{T}_1 = \begin{pmatrix} c_\psi^2 & -s_\psi c_\psi^2 & 0 \\ s_\psi c_\psi & s_\psi^2  & 0 \\ 0& 0& 0\end{pmatrix}, \nonumber\\
\mathcal{T}_2 = \begin{pmatrix} s_\psi^2 s_\alpha^2 & c_\psi s_\psi s_\alpha^2 & -s_\psi c_\alpha s_\alpha \\  
c_\psi s_\psi s_\alpha^2 & c_\psi ^2 s_\alpha^2 & -c_\psi c_\alpha s_\alpha \\
-s_\psi c_\alpha s_\alpha & -c_\psi c_\alpha s_\alpha &  c_\alpha^2 \end{pmatrix},&
\mathcal{T}_3 = \begin{pmatrix} s_\psi^2 c_\alpha^2 & c_\psi s_\psi c_\alpha^2 & s_\psi c_\alpha s_\alpha \\  
c_\psi s_\psi c_\alpha^2 & c_\psi ^2 c_\alpha^2 & c_\psi c_\alpha s_\alpha \\
s_\psi c_\alpha s_\alpha & c_\psi c_\alpha s_\alpha &  s_\alpha^2 \end{pmatrix}.
\end{eqnarray}
For $N$ consecutive domains with angle $\psi_n$ in each domain,
 it can be shown that the total transfer matrix is given as a product over all domains,
\begin{equation}
\mathcal T(x_{3,N},\ldots,x_{3,1};\psi_N,\dots,\psi_1;E) = \prod\limits_{n= 1}^N \mathcal T(x_{3,n+1},x_{3,i};\psi_n; E).
\label{eq:prod}
\end{equation}
Present $\gamma$-ray experiments cannot measure the polarisation. Therefore, one has to generalise the problem at hand 
to the density matrix formalism, where $\rho = \Psi \otimes \Psi^\dagger$~\cite{mirizzi2009}. The probability for an initially unpolarised photon beam,
$\rho_\mathrm{unpol} = 1/2\mathrm{diag}(1,1,0)$, to oscillate into an ALP, $\rho_{aa} = \mathrm{diag}(0,0,1)$ is then given by~\cite{mirizzi2009}
\begin{equation}
P_{a\gamma} = \mathrm{Tr}\left(\rho_{aa}\mathcal T \rho_\mathrm{unpol} \mathcal T^\dagger\right).\label{eq:pag}
\end{equation}
The oscillation probability will take some value $0 \leqslant P_{a\gamma} \leqslant 1/2$~\cite{meyer2013thesis} depending on the realisation of the angles $\{\psi_n\}$ and the mixing angle. 
Interestingly, realisations exist for which $P_{a\gamma} = 0$ even though $\alpha > 0$. 
To show this we assume an even number of domains where $\psi = c\pi$ in one half of the domains and $(c + 1)\pi$  in the other half (where $c$ is a real non-zero number)
ordered randomly. 
A straightforward calculations shows that the commutator of the transfer matrices $C = [\mathcal T(\psi = (c + 1)\pi),\mathcal T(\psi = c\pi)]$ is an anti-symmetric matrix 
with entries
\begin{eqnarray}
C_{13} = -C_{31}  &=& \frac{1}{2} \left(e^{i \lambda_2x_3} - e^{i \lambda_3x_3}\right)^2 s_{c \pi}s_{4 \alpha}\\
C_{23} = -C_{32 } &=& \frac{1}{2} \left(e^{i \lambda_2x_3} - e^{i \lambda_3x_3}\right)^2 c_{c \pi}s_{4 \alpha}
\end{eqnarray}
and zero  in all other entries. The matrix elements of the product $\mathcal T(\psi = (c + 1)\pi)\mathcal T(\psi = c\pi)$
that induce mixing (i.e., the $i3$, $i = 1,2$ elements in the current basis) are found to be equal to $2C_{i3}$.
Above a critical energy $E_\mathrm{crit}$  the mixing becomes independent of energy. If in addition the mixing is strong so that $\alpha \to \pi / 4$, the commutator and the mixing inducing matrix elements vanish. 
With the commutator equal to zero we can now combine all pairs of $c\pi$ and $(c+1)\pi$ transfer matrices and see that the resulting product 
of all matrices given in Eq. \eqref{eq:prod} does not induce any photon-ALP mixing. 

\begin{figure}[t]
\centering
\includegraphics[width = .7\linewidth]{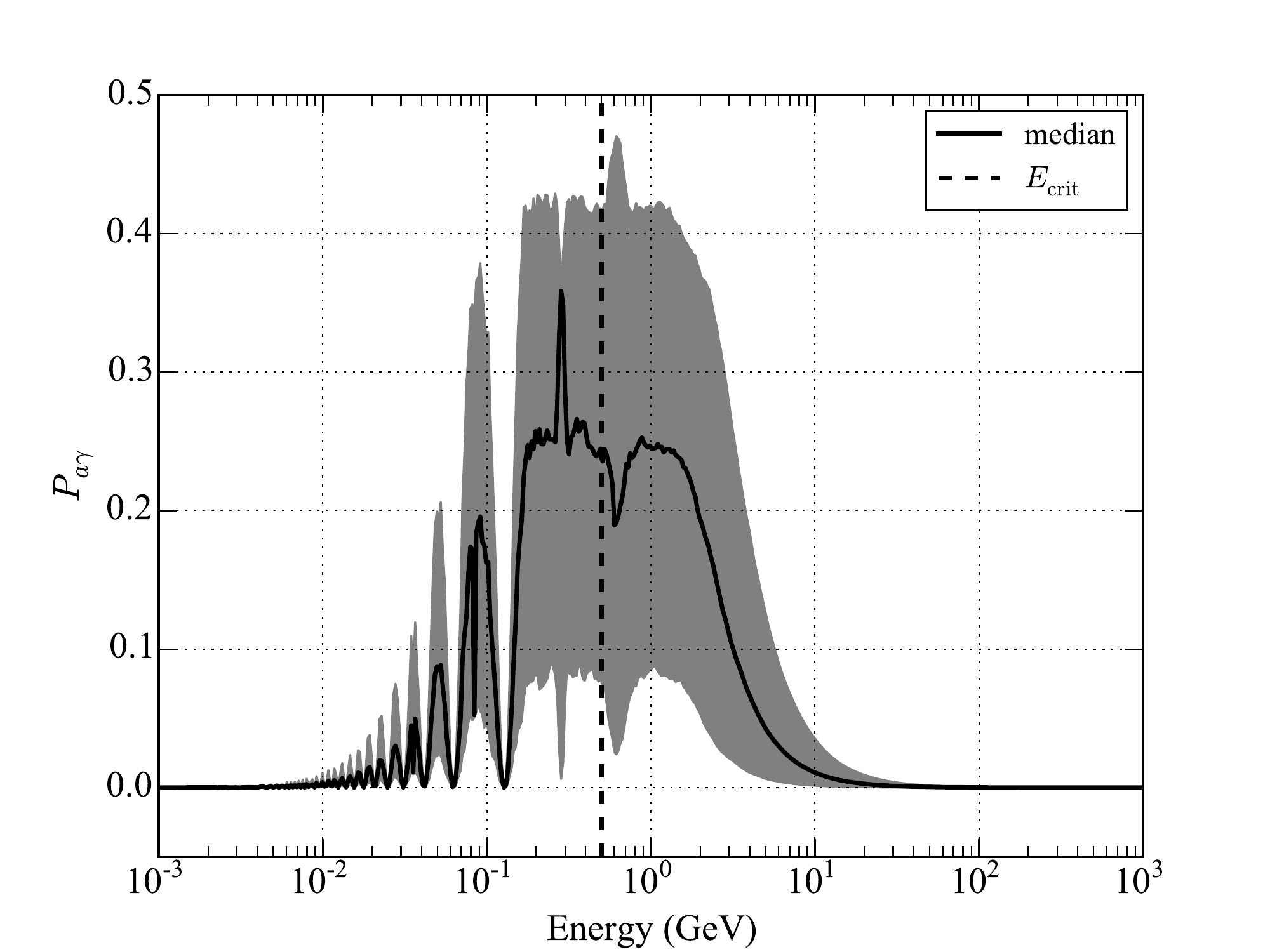}
\caption{Conversion probability in 30 domains with $L_\mathrm{coh} = 10\,$kpc and $B = 1\,\mu$G. Half of the domains have $\psi = 0$ 
while in the other half $\psi = \pi$ is chosen. The solid line shows the median of $P_{a\gamma}$  for 1000 random permutations of the angles. The shaded area gives the probability in the 68\,\% interval around the median. The dashed line shows $E_\mathrm{crit}$ above which $\alpha \to \pi/4$.
An ALP mass of 1\,neV and a photon-ALP coupling of $5\times10^{-11}\,\mathrm{GeV}^{-1}$ are assumed.}
\label{fig:ex}
\end{figure}

As an example, we show this behaviour in Fig. \ref{fig:ex}, in which we assume magnetic-field parameters found in galaxy clusters.
 The conversion probability is calculated numerically following Eq. \eqref{eq:pag}. Above the critical energy the
probability goes to zero, however, around the critical energy oscillations still occur. 
Our findings still hold even if photon absorption is included as it is the case for conversions in the intergalactic magnetic field. 
However, as this magnetic field evolves with redshift, not all realisations lead to a conversion probability exactly equal to zero for all random permutations. 

\section{Conclusions}
As shown in the previous section, the photon-ALP conversion probability can be exactly zero in special configurations 
of a turbulent magnetic field given that (a) it is modelled with a simple cell-like structure and (b) that the mixing
occurs in the strong mixing regime, i.e. at energies $> E_\mathrm{crit}$ and $\alpha \to \pi / 4$. Oscillations around the critical energy still occur
making spectral features at this energy a universal prediction of photon-ALP oscillations.
The absence of such signatures in $\gamma$-ray spectra has already been used to constrain the photon-ALP coupling~\cite{hess2013:alps}. 
In more realistic models of the turbulent field (that use, e.g., a Kolmogorov turbulence spectrum) 
we do not have the freedom to choose the $\psi$ angles (see, e.g. Ref.~\cite{meyer2014}) and we cannot easily construct a scenario with vanishing mixing as done here.
Utilizing such models, it can be shown that the future Cherenkov Telescope Array will be sensitive to detect a boost in 
the photon flux for photon-ALP couplings $\gtrsim 2\times10^{-11}\,\mathrm{GeV}^{-1}$ and ALP masses $\lesssim 100\,\mathrm{neV}$
\cite{meyer2014cta}, the same parameters that could explain evidence for a reduced opacity for VHE $\gamma$-rays~\cite{meyer2013}.

\section*{Acknowledgments}
MM is supported by a grant of the Knut and Alice Wallenberg Foundation, PI: Jan Conrad
 

\begin{footnotesize}


\end{footnotesize}


\end{document}